\documentclass[preprint]{revtex4}
\usepackage[latin1] {inputenc}  
\usepackage{graphicx}  
\usepackage{color}
\usepackage{amsmath}   
\usepackage{array}
\usepackage{amsfonts} 
\usepackage{amssymb}
\usepackage{epsfig}
\usepackage{theorem}
\usepackage{boxedminipage}
\usepackage{pdfsync}
\ifx \theorem \undefined
\newtheorem{definition}{\vspace{1mm}Definition}[section]

\newtheorem{example}[definition]{\vspace{1mm}Example}
\newtheorem{theorem}[definition]{\vspace{1mm}Theorem}

\newtheorem{corollary}{\vspace{1mm}Corollary}
\newtheorem{proposition}[definition]{\vspace{1mm}Proposition}

\fi

\newenvironment{proof}{\begin{trivlist}\item{\bf Proof.}}{\end{trivlist}}

\newcommand{\qu}[1]{\ensuremath{|#1\rangle}}
\newcommand{\dqu}[1]{\ensuremath{\langle #1|}}

\newcommand{\ign}[1]{}

\newcommand{\nats}{\ensuremath{\mathbb{N}}}

\begin{document}
\title{Improving Classical Authentication with Quantum Communication  } 

\author{F.\ M.\ Assis}
\affiliation{Departament of Electrical Engineering \\
 UFCG, Universidade Federal de Campina Grande,  Brazil}

\author{P. Mateus}
\affiliation{IST, Technical University of Lisbon, P-1040-001 Lisbon, and
SQIG, Instituto de Telecomunica\c{c}\~oes, P-1049-001 Lisbon, Portugal} 

\author{Y. Omar}
\affiliation{CEMAPRE, ISEG, Universidade T\'{e}cnica de Lisboa, P-1200-781 Lisbon, and
SQIG, Instituto de Telecomunica\c{c}\~oes, P-1049-001 Lisbon, Portugal}

\date{11 February 2010} 

\begin{abstract}
We propose a quantum-enhanced protocol to authenticate classical messages, with improved security with
respect to the classical scheme introduced by Brassard in 1983. In that protocol, the shared key is the
seed of a pseudo-random generator (PRG) and a hash function is used to create the authentication tag of a
public message. We show that a quantum encoding of secret bits offers more security than the classical XOR
function introduced by Brassard. Furthermore, we establish the relationship between the bias of a PRG and
the amount of information about the key that the attacker can retrieve from a block of authenticated
messages.  Finally, we prove that quantum resources can improve both the secrecy of the key generated by 
the PRG  and the secrecy of the tag obtained with a hidden hash function.  
\end{abstract}

\maketitle

\section{Introduction}

The authentication of public messages is a fundamental problem
nowadays for bipartite and network communications.
The scenario is the following: Alice sends a (classical)
message to Bob through a public channel, together with an
authentication tag through a private or public channel. The tag will allow
Bob to verify if the message he received via the public channel
has been tampered with or if it is indeed the authentic message,
originally sent by Alice. A third character, Eve, wants to
sabotage this scheme by intercepting Alice's message and sending
her own message to Bob, together with a false tag which will
convince Bob he is receiving the authentic message. For instance,
one could imagine that Alice is sending to Bob her bank account
number, to which Bob will transfer some money, and Eve wants to
interfere in the communication in such a way that Bob will receive
her bank account number believing it is Alice's one, thus giving
his money to Eve.  The use of authentication tags allows to 
separate the secrecy problem in message transmission from the authentication problem 
and it is useful even if a secure communication channel is 
available~\cite{WeCa:81}. 
 
In 1983, G. Brassard proposed a computationally secure scheme of
classical authentication tags based on the sharing of short secret
keys~\cite{Bras:83}.  Brassard's scheme is itself an 
improvement of the Wegman-Carter protocol~\cite{WeCa:81}. Brassard showed 
that a  relatively short seed of a PRG can be used as a secret key shared 
between Alice and Bob which will allow the exchange of computationally secure 
authentication tags. This method yields a much more practical protocol, where 
the requirements on the  seed length grow reasonably with the number of messages 
we want to authenticate, as  opposed to the Wegman-Carter proposal. 

The security of PRGs is based on the alleged hardness of some problems 
of number theory, e.g., the factorization of a large number with classical
computers. However, several of these  problems are provably solvable if 
quantum computers are available.  
Consequently, the security of the PRGs might be compromised.
Assuming Alice and Bob communicate quantically, can Eve yet menacing the PRG
security? This question is our main motivation to write this article. 

In this work, we extend Brassard's protocol to include quantum-encoded 
authentication tags, which we prove will offer, under certain conditions, 
information-theoretical security for the authentication of classical messages.  
We observe that our scheme can authenticate the quantum channel itself, which
is an important part of the quantum cryptography: in fact, it is the crucial first step of quantum key distribution
protocols.

\section{\label{sec:preliminares} Preliminaries}
In this section we set up basic notation, briefly review the description 
of the Brassard's protocol and describe our new  proposal. We conclude the section
with a negative result on the robustness of an attackable PRG when its output is 
hidden by a specific quantum coding.  

\begin{figure}[htbp]
\centerline{\epsfig{file=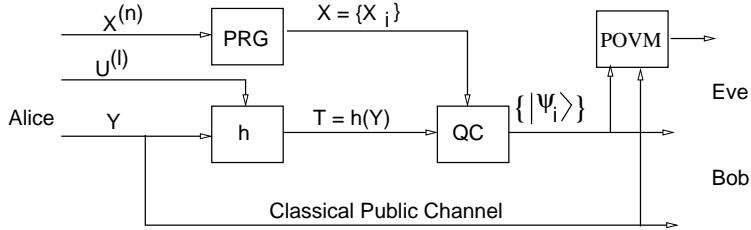,scale=0.5}}
\caption{First proposal of quantum-enhanced authentication scheme  }
\label{f:qauthproposed}
\end{figure}

We denote $\mathcal{M} $ the set of messages and $\mathcal{T} $ the set of tags, 
where $\log |\mathcal{M} | >> \log |\mathcal{T}| $.
As hash functions are an important ingredient for all protocols  described here 
we start by presenting their formal definition~\cite{larsson08.1}: 
\begin{definition}[$\varepsilon-\textrm{almost}$ strongly universal-2 hash functions]
\ \ Let $\mathcal{M} $ and $\mathcal{T} $ be finite sets and call functions 
from $\mathcal{M} $ to $\mathcal{T} $ \textit{hash functions}. Let $\varepsilon $ be a
positive real number. A set $\mathcal{H} $ of hash functions is $\varepsilon-$almost strongly
universal-2 if the following two conditions are satisfied
\begin{itemize}
\item[1) ] The number of hash functions in $\mathcal{H} $ that takes an arbitrary $m\in
\mathcal{M} $ to an arbitrary $t \in \mathcal{T} $ is exactly $|\mathcal{H}| /|\mathcal{T}|.$
\item[2) ] The fraction of those functions that also takes $Y^{\prime} \neq Y $ in 
$\mathcal{M} $ to an arbitrary $T^{\prime} \in \mathcal{T} $ (possibly equal to $T$) is no
more than $\varepsilon .$
\end{itemize}
\end{definition}
The number $\varepsilon $ is related to the probability of guessing the correct tag with respect to an arbitrary message $Y$.  
Notice that the smaller $\varepsilon$ is, the 
larger is $|\mathcal{H} | $. 
\begin{figure}[htbp]
\centerline{\epsfig{file=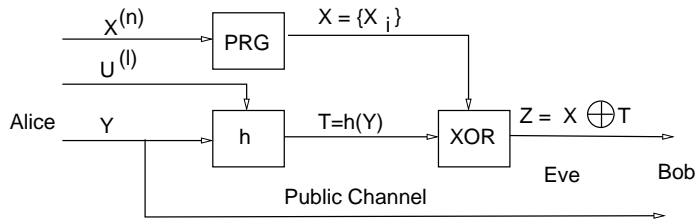,scale=0.5}}
\caption{Brassard's classical authentication protocol~\cite{Bras:83}  }
\label{f:qauthbrassard}
\end{figure}
For additional details on universal-2 functions we point the reader to~\cite{WeCa:81}.
Brassard's protocol (see Figure~\ref{f:qauthbrassard})  makes use of two 
secret keys. The first one, $U^{(l)},$
specifies a fixed universal-2 hash function $h \in {\cal H},$ where  
$l = \lceil \log |\mathcal{H} | \rceil .$  
The second specifies the seed $X^{(n)}\in \mathbb{Z}_2^n $, 
for a PRG, a sequence of $n$ bits. 
The main ingredient of our first quantum-enhanced protocol proposed here
(see Figure~\ref{f:qauthproposed})  is replacing the classical  gate \texttt{XOR} of
Brassard protocol by a quantum encoder 
similar to that used in the BB84 protocol~\cite{BB84:84}. After some 
developments,  we shall  verify that the key $U^{(l)} $ is no longer necessary.  
Assume that Alice and Bob agree on two orthonormal bases $B_0$ and $B_1$ for the 
2-dimensional Hilbert space,
$$
B_0 = \left\{\qu{0},\qu{1} \right\} \ \ \text{ and }
B_1 = \left\{\qu{+} = \frac{1}{\sqrt{2}} (\qu{0}+\qu{1}),
      \qu{-} = \frac{1}{\sqrt{2}} (\qu{0}-\qu{1}) \right\}
$$
These bases will be used to prepare four quantum states. We shall refer
to this preparation process as \textit{quantum coding}. For each bit of the
$ k = \lceil \log |\mathcal{T} | \rceil $ bits long tag $ T_Y = h(Y) $,
Alice prepares a quantum state  $\qu{\psi} = \qu{\psi }\left(X_i, (T_Y)_i\right)  $ 
determined by 
the bit $X_i$ from the PRG and the corresponding bit $(T_Y)_i $ of 2-radix representation 
of the tag $T_Y. $  Then, if the bit $X_i = 0, $ Alice prepares $\qu{\psi}$ using 
basis $B_0$, such that
\begin{equation} \label{eq:B0}
\qu{\psi}=
\begin{cases}
    \qu{0} & \quad \text{ if } (T_Y)_i =  0 \\
    \qu{1} & \quad  \text{if } (T_Y)_i =  1.
\end{cases}
\end{equation}
Similarly, if the bit $X_i = 1 $, Alice prepares $\qu{\psi}$ using basis  $B_1$,
such  that
\begin{equation} \label{eq:B1}
\qu{\psi}=
\begin{cases}
    \qu{+} & \quad \text{ if } (T_Y)_i = 0 \\
    \qu{-} & \quad \text{ if } (T_Y)_i = 1
\end{cases}
\end{equation}
After the qubits generation, Alice sends the separable state 
$\qu{\psi_{Y}}^{\otimes k}$ to Bob through a noiseless quantum channel and the message 
$Y$ through an unauthenticated classical channel. At the reception, Bob performs measurements 
to obtain  a sequence of $k$ bits from the quantum encoded version of $h(Y).$  For the 
$i$-th received qubit, Bob measures it using the basis $ B_0$ or $B_1$ depending on the 
$i-$th bit of $X$ is 0 or 1, respectively, recovering a $k$-bit long string  
$T^{\prime} = h^{\prime}\left(\qu{\psi}^{\otimes k}\right) $.

Because the quantum channel is assumed to be perfect, Bob recognizes that the message is 
authentic if $h^{\prime} = h(Y_B)$, where $Y_B$ is the message received from the classical
channel. Otherwise, Bob assumes that Eve tried to send him an unauthentic message. This concludes the authentication protocol for one message. Throughout this article it is always assumed that the above
coding rule is public.  

Even though we assume a noise-free quantum channel, we observe that if the 
quantum channel is noisy, the only piece of information requiring 
error-protecting coding is the block of bits $(T_Y)_i $ of the tag $T_Y $.
The sequence of bases to be prepared by Alice and Bob is known a priori, determined
locally by the sequence of bits from the PRG. 
A future task is evaluating the effects of the utilization of error-correcting codes
to the bits of $T_Y$. 

In a warning against alleged collective attacks, we notice that our analysis allows
Eve to make general procedures (suggested in Figure~\ref{f:qauthproposed} by the block 
labeled POVM) without being detected. Our results are robust to such powerful and unrealistic
 assumption for the attacker. Note that our quantum scheme aims at minimizing the key length for
one-way transmission. Another example of such an approach is given in~\cite{Damgaard:04}. Next we 
focus on crucial aspects of the PRGs.
\subsection*{Weak pseudo-random generators } 
Clearly, it is important to understand how secure the authentication code described 
above is. As we shall see, the security of the authentication code is deeply related 
with to quality of the pseudo-random generator.  The quality of a pseudo-random generator 
is evaluated  by  the hardness to discriminate its pseudo-random sequence output from 
a truly random sequence or by the hardness to find its seed. The first quality 
evaluation relates to the PRG's robustness against \textit{distinguishing attacks}, the 
second  relates to the so-called \textit{state recovery attacks}. In~\cite{sidok05.1} 
it is shown that a state recovery attack is a subclass of the distinguishing attacks.

\begin{figure}[htbp]
\centerline{\epsfig{file=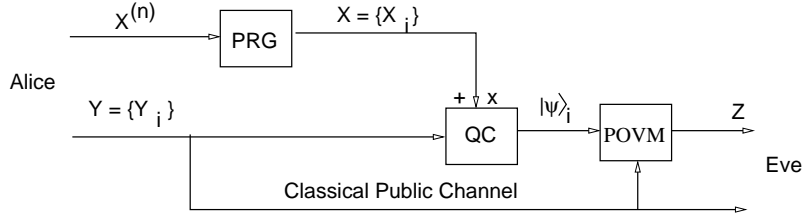,scale=0.5}}
\caption{Auxiliary scheme for Theorem~\ref{thm:negativo} proof }
\label{f:qauth3}
\end{figure}

As a matter of fact, if the pseudo-random generator can be
attacked by a quantum computer so does the authentication code. To set this result we
refer to Figure~\ref{f:qauth3}, that describes a simple scheme to assist us  the proof. 
In this scheme, we simply allow Eve to compare a sequence $\{Y_i \} $ of classical bits 
with the corresponding sequence $\{ Z_i \} $ obtained from the measurement apparatus 
POVM. 

Recall that a pseudo-random generator is a polynomial-time family
of functions $G=\{G_n:\mathbb{Z}_2^n\times\nats\to \mathbb{Z}_2\}_{n\in \nats}$ where 
$\mathbb{Z}_2$ is  the set $\{0,1\}$ and $G_n$ is the pseudo-generator for seeds with size 
$n$, that is, $G_n(X^{(n)}, i)$ returns the $i$-th bit generated from $n$ bits long seed 
$X^{(n)}$. Pseudo-random generators are expected to fulfill an indistinguishability property 
that we will not detail here for the sake of simplicity (more details on~\cite{goldreich99}).
In the following definition we write
$
X^{p(n)}  = \left( G(X^{(n)},i_1), G(X^{(n)}, i_2), \ldots , G(X^{(n)}, i_{p(n)}) \right)
$
to denote a subsequence of $p(n)$ (not necessarily contiguous) bits generated 
by $G.$

\begin{definition}\em We say that a pseudo-random generator $G$ is
{\em attackable in (quan\-tum/proba\-bi\-listic) polynomial time} if there exists a
(quantum/probabilistic) polynomial time algorithm $P$ and polynomial $p$ such that if 
$P$ is fed with a subsequence of $p(n)$ (not necessarily contiguous) generated bits 
$X^{p(n)}$ of $G$ we have that:  
                 \[ H(X^{(n)}|P(X^{p(n)}))\in O(2^{-n}).\] 
\end{definition}
For a pseudo-random generator to be attackable, there must exist an algorithm (quantum 
or probabilistic) that receives a subsequence of $p(n)$ generated bits (not necessarily
contiguous) and is able to compute the seed up to a negligible uncertainty. We observe
that the security/randomness of the pseudo-random generator can not be grounded in the 
fact that the attack can only be performed to a contiguous subsequence of generated bits. 
This is due to the fact that the generator could always hide some bits if the attack 
required this type of sequences.
A simple example of a pseudo-random generator that can be attackable in polynomial time 
are  the pseudo-number generators based on linear congruence~\cite{sidok05.1}.

\begin{theorem}\label{thm:negativo}\em If a pseudo-random generator $G$ is
{\em attackable in (quantum/proba\-bilistic) polynomial time} then the scheme presented 
in Figure~\ref{f:qauth3} is not secure in polynomial-time for a quantum adversary that 
has access to $Y = \{ Y_i\} $.
\end{theorem}
\begin{proof} Since $G$ is attackable there exists a quantum polynomial time algorithm 
$P$ and a polynomial $p$ such that if $P$ is fed with $p(n)$ bits of the string $X$ 
generated by $G$ then $P$ computes (up to negligible uncertainty) the seed $X^{(n)}$ of 
$G$. So it is enough to show that Eve, upon capturing the qubits generated by 
$\texttt{QC}$, is able to recover (with non-negligible probability) $p(n)$ bits of $X$.

Indeed, assume that Eve has captured $8 p(n)$ qubits $\qu{\psi}_i, i:1\dots 8 p(n)$ and 
has measured them in a random basis (that is, either the computational or the diagonal 
basis). Eve can now verify if $Z_i = Y_i . $  If this  occurs  Eve does not now if the 
basis chose to encode the $Y_i$ bit was the basis she measured or if she got with 
$\frac{1}{2}$ probability  the correct bit due to encoding in the other basis. However, 
if the outcome is different (that is, $Y_i \neq Z_i $ ), then she knows that the 
basis at the $i-th$ bit is the basis she did not choose the measure, because no mismatch 
would be possible if the encoding  was performed with the same basis. In the latter 
case, she knows that $X_i $ is either $0$ or $1$ depending if she measured in the 
diagonal or the computational basis,  respectively. Moreover, this happens
with $1/4$ probability. So the probability of Eve not obtaining $p(n)$ elements of $X$ 
by measuring $8p(n)$ qubits is given by the cumulative function of a binomial 
distribution with $1/4$ Bernoulli trial, $8p(n)$ trials and success of at most $p(n)$. 
By Hoeffding's inequality this probability is upper-bounded by
$\exp\left(-2\frac{(8p(n)/4-p(n))^2}{p(n)})\right)=\exp(-2p(n))$ which decreases
exponentially with $n$, and so in other words, Eve has an exponentially increasing
probability of obtaining $p(n)$ bits of $X$ with $8p(n)$ qubits measurements. Since $G$ 
is attackable by knowing $p(n)$ bits of $X$, Eve is able to perform this attack up to
negligible probability. \hfill$ \square $ 
\end{proof}

\begin{corollary}
\label{corollary:1}\em 
If a pseudo-random generator $G$ is attackable then the scheme presented in 
Figure~\ref{f:qauthproposed} is not secure in polynomial-time for a quantum
adversary that has access to hash function $h.$
\end{corollary}

\begin{proof}
Eve is able to  calculate  $h(Y)$ from $Y$ that is  public. Therefore she can
apply Theorem~\ref{thm:negativo} by observing a number $N$ of tags such that 
$N \log|\mathcal{T} | \geq 8p(n) .$  \hfill $\square $ 
\end{proof}

Although Theorem~\ref{thm:negativo} points that the quantum coding of
Figure~\ref{f:qauth3} is not better asymptotically than the classical coding  (where
we simply replace the quantum coder $\texttt{QC}$ by a $\texttt{XOR} $ gate),  it seems 
harder to attack the quantum scheme. We will now show that this is true for the 
simple case where the encoder is fed by an independent and identically distributed 
(i.i.d.) Bernoulli sequence. The following example illustrates that this is true even for 
a very simple generator.
\begin{example}[State Recovery Attack for Linear Congruential Generator(LCG)] \em
Let $A$ be a positive integer and $\mathbb{Z}_A $ 
the set of integers modulo $A.$  The seed of the LCG is the vector 
$X^{(n)} = (A, s_0, a, b )$,  where $s_0, a, b \in \mathbb{Z}_A $. The length of the 
seed is $n = 4\lceil \log A \rceil $.
A binary pseudorandom sequence with length $N\times \lceil \log A \rceil $ bits is 
obtained from the 2-radix expansion of the sequence 
$\mathbf{s} = \{ s_1,\ s_2, \ldots , s_N \} $  created by the following recursion:
\begin{equation}\label{e:weakLCG}
s_i = a s_{i-1} + b \mod A, \ \ i=2,3, \ldots , N 
\end{equation}
It is well known (see~\cite{sidok05.1}) that for all $i, i=1,2, \ldots, N-3$, the numbers 
\[
\delta_i = \mathrm{det} \left[\begin{array}{lll} s_i & s_{i+1} & 1 \\ 
                         s_{i+1} & s_{i+2} & 1 \\ s_{i+2} & s_{i+3} & 1 
                        \end{array}\right]
\]
are multiple of $A$. As a consequence, the greatest common divisor $\mathrm{GCD} $ of some 
$\delta_i's$ gives the value of $A$. The rest of the seed, that is $a, b $ and $s_0 $, 
follow then from a system of linear  equations. In practice five values of $\delta_i $ 
are enough. 

Figure~\ref{f:cqbits}(right) displays a simplified version of the scheme shown in 
Figure~\ref{f:qauth3}, where $X$ stands for the pseudo-random sequence from the output of
the PRG. The left side of Figure~\ref{f:cqbits} displays the situation when  a
gate \texttt{XOR} is utilized. We notice that the state recovery attack is applicable 
without  change to the \texttt{XOR}-based scheme. It is enough to compute 
$X = Z \oplus Y $ before  applying the algorithm.  
In contrast, for the quantum scheme, Eve is submitted to an irreducible uncertainty 
on the $X$ values due to quantum coding. In particular, if she employs the procedure 
described in the proof of the Theorem~\ref{thm:negativo} it is expected only one fourth
of the $X$'s are expected to be correct. The problem from  Eve's point of view is how to solve the
seed from a degraded version of the algorithm input $X$. 
\end{example}

\section{\label{sec:comparing} Comparing \texttt{XOR} with quantum coding}
In the last section we have considered the problem of the state recovering attack and defined
the weakness of a PRG. In this section we make a rigorous comparison between the \texttt{XOR} and the quantum coding performances using information-theoretical measures. 
\begin{figure}[htbp]
\centerline{\epsfig{file=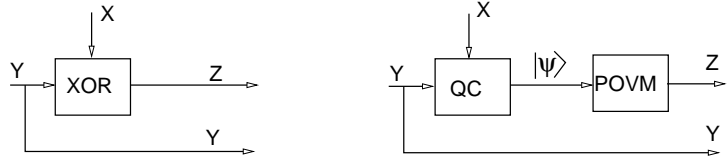,scale=0.5}}
\caption{\texttt{XOR} (left) and quantum coding (right) }\label{f:cqbits}
\end{figure}
To this end consider Figure~\ref{f:cqbits} where both classical and quantum encodings are displayed. The
$\texttt{QC}$ denotes the 
quantum encoder defined before, in \eqref{eq:B0} and \eqref{eq:B1}, where $X$ is the  variable that sets the basis.
 The block POVM  stands for the measurement apparatus defined by the positive operator-valued measure
$$Z=\{E_m(Y)\}_{m\in O}$$ where $O$ is the set of outcomes. Observe that the measurement  may depend on 
the message $Y$, which is public.
The goal of Eve is to maximize the knowledge of $X$, that is, minimize the entropy $H(X|Y,Z)$.

We consider the classical and quantum scheme presented in Figure~\ref{f:cqbits} in two ways: 
Firstly, we will assume that $X$ is a sequence of fair and independent Bernoulli random variables, 
that is, the PRG describing $X$ is perfect. Secondly, we consider a biased PRG (unfair) to describe $X$ and  introduce blocks of random variables into the analysis.

\subsection*{Fair input single-sized block}
We start with the simple case of a single-sized block and where $X\sim\textrm{Ber}\left(\frac{1}{2}\right)$. In the classical \texttt{XOR} encoding case we have that $Z=X\oplus Y$ and thus $H(X|Y,Z)=0$, and so Eve has no doubt about $X$.
 In the quantum encoding case,  the Holevo bound states that
\begin{equation}\label{ineq:1}
I(X;Z|Y)\leq S(\rho(Y)) -\sum_{i=0}^1\frac{1}{2}S(\qu{\phi_i(Y)}\dqu{\phi_i(Y)})
\end{equation}
where $\rho(Y)$ is the density operator describing the encoding by $\textrm{QC}$ that is
\begin{equation}
	\rho(Y)= \frac{1}{2}\qu{\phi_0(Y)}\dqu{\phi_0(Y)} + \frac{1}{2}\qu{\phi_1(Y)}\dqu{\phi_1(Y)},
\end{equation}
where  $\qu{\phi_0(0)} = \qu{0}$, $\qu{\phi_1(0)} = \qu{+}$, 
$\qu{\phi_0(1)} = \qu{1}$ and  $\qu{\phi_1(1)}=\qu{-}$.

We shall need a well known property of the von Neumann entropy~ (see \cite{NC:2000} for more details). 
\begin{proposition}\label{zeroS}\em 
Let $\mathbf{\rho} $ be a quantum state and $S(\mathbf{\rho} ) $ its entropy,  then  
$ S(\mathbf{\rho} ) \geq 0 $,  and the 
equality holds iff  $\rho $  is a pure state. 
\end{proposition}
 Thus, thanks to Proposition~\ref{zeroS} we can simplify \eqref{ineq:1} to
\begin{equation}\label{ineq:2}
I(X;Z|Y)\leq S(\rho(Y)).
\end{equation}
Moreover, one can compute easily the von Neumann entropy of $S(\rho(Y))=S(\rho(0))=S(\rho(1))$ and is
\begin{eqnarray}\label{e:xyztheta1}
S^{\ast}=S(\rho(Y))
  & = &  -2 \cos^2\left(\frac{\pi }{8}\right) \log\left(\cos\left(\frac{\pi }{8}\right)\right)-2 \log\left(\sin\left(\frac{\pi }{8}\right)\right) \sin^2\left(\frac{\pi }{8}\right).
\end{eqnarray}
And so, since $H(X|Z,Y)=H(X|Y)-I(X;Z|Y)$ and $H(X|Y)=1$, the minimum uncertainty that Eve may attain about
 $X$ is given by
\begin{eqnarray}\label{e:xyztheta}
H\left(X| Y,Z \right)
  & = &     1-S(\rho(Y)).
\end{eqnarray}
The Holevo bound can be achieved by a simple von Neumann measurement~\cite[pp.421]{Paris:04} described 
by the Hermitian
\begin{equation}
	A=\mathbf{0}\qu{\psi_\theta}\dqu{\psi_\theta} + \mathbf{1}\qu{\psi^\bot_\theta}
	               \dqu{\psi^\bot_\theta}
\end{equation}
with $\psi_\theta=\cos(\theta)\qu{0}+\sin(\theta)\qu{1}$,
$\psi^\bot_\theta=\sin(\theta)\qu{0}-\cos(\theta)\qu{1}$ and
$\theta=-\frac{\pi}{8}$.

\subsection*{Fair input $k$-blocks}

First, consider the classical setup, then $H(X^k| Y^k, Z^k ) = 0$ , since the block $X^{k}$ is 
completely determined from the knowledge of $Y^k$ and $Z^k$.  

For the quantum setup, the subsystem that Eve owns is 
described by  
\begin{equation}\label{fairk}
	\rho_{Y^k} = \bigotimes_{i=1}^k \left(\frac{1}{2}\qu{\phi_0(Y_i)}\dqu{\phi_0(Y_i)} + \frac{1}{2}\qu{\phi_1(Y_i)}\dqu{\phi_1(Y_i)}\right). 	
\end{equation}

By the Holevo bound we get that
\begin{equation}\label{e:hs2} 
H\left( X^k|Y^k, Z^k \right)  \geq H(X^k)-  S(\rho_{Y^k} ).
\end{equation}

\begin{example}\em Table~\ref{tab:t1} illustrates the scenario for $k=2.$ Rows are indexed by the
four possible values of $Y^2$ and columns are indexed by the bases 
corresponding to the four values of $X^2.$ Notice that Eve is not able 
to distinguish which column is being used. Then, her uncertainty is lower bounded by the  
von Neumann entropy of the quantum system formed by states listed in
row indexed by the values of $Y^2 $ that she can access.  
\begin{table}[hbtp] 
\caption{Encoding for blocks of length 2  \label{tab:t1}}
\begin{center}
\begin{tabular}{l|l|l|l|l}
\hline \hline
& \multicolumn{4}{c}{Bases }  \\ \cline{2-5}
\multicolumn{1}{c|}{$Y^2$} & $B_0B_0$ & $B_0B_1$ & $B_1B_0$ & $B_1B_1$   \\ \hline
$00$ & $\qu{00} $ & $ \qu{0+}$  & $ \qu{+0} $ & $ \qu{++} $  \\ \hline
$01$ & $\qu{01} $ & $ \qu{0-} $ & $ \qu{+1} $ & $ \qu{+-} $  \\ \hline
$10$ & $\qu{10} $ & $ \qu{1+} $ & $ \qu{-0} $ & $ \qu{-+} $  \\ \hline
$11$ & $\qu{11} $ & $ \qu{1-} $ & $ \qu{-1} $ & $ \qu{--} $  \\ \hline
\end{tabular}
\end{center}
\end{table}

\end{example}

Recall the following property concerning the von Neumann entropy.

\begin{proposition}\label{2prop}\em Let $\rho$ and $\sigma$ be quantum states, then
$S\left( \mathbf{\rho} \otimes \mathbf{\sigma} \right) = S(\mathbf{\rho} ) 
                                                       + S(\mathbf{\sigma} ).$ 
\end{proposition}
As a consequence of Equation \eqref{fairk} and Proposition~\ref{2prop}, for a sequence of fair Bernoullis we have
\begin{equation}\label{e:somasmax}
S(\rho_{Y^k} ) = kS^{\ast},
\end{equation}
where $S^{\ast}$ is given by \eqref{e:xyztheta}. So we have that 
\begin{equation}\label{e:hs2b} 
H\left( X^k|Y^k, Z^k \right)  \geq k -  kS^{\ast}.
\end{equation}
Again, the equality can be achieved by a simple von Neumann measurement, namely that defined by $A^{\otimes k}$. This is the best scenario one can imagine to defeat Eve. However, for the protocol to be practical, the $X's$ should be generated by a PRG, which is the case we examine next.
\subsection*{Unfair input $k$-blocks}
The results above where obtained assuming that $\{ X_i \} $
was a sequence of i.i.d. fair Bernoulli random variables. In this section we study the general case, 
with the purpose of clarifying how  the use of a real PRG affects the uncertainty about $X$.


Consider $k-$length blocks $X^{k},\ Y^{k} $ and $Z^{k} ,$ where 
$X^{k} =  X_{i+1}, X_{i+2}, \ldots , X_{i + k} $   is a 
contiguous subsequence of $ \{X_i\} $ and similarly to $Y^{k} $ and $Z^{k}. $
Note that, to ease notation,  we omit the index $i$ in defining $X^k.$ However, it is 
crucial to remark that the probability distribution of $X^k$ is, in general, dependent 
on $i$. As a matter of fact, $\mathbf{p}_{X^k} = \left( p_0, p_2,\ldots , p_{2^k-1} \right) $ 
can even degenerate to a distribution with a single component equal to 1,  depending on the 
robustness of the  PRG. We shall simplify the notation denoting $\mathbf{p}_{X^k}$ by
$\mathbf{p} $.

Concerning the unfairness of $\{X_i\}$, the best strategy for Eve to get information from 
$X^k$ is to prepare a measurement (POVM) over the all $k$ qubits sent, given that she knows  $Y^k$.  Again, the Holevo bound gives us


\begin{equation}\label{e:hs4} 
H\left( X^k|Y^k, Z^k \right)  \geq H(X^k)-  S(\rho_{Y^k} ) = 
H(X^k)-H(\lambda ) 
\end{equation}
where $\lambda=(\lambda_1\dots\lambda_{2^k})$ is the spectrum of $\rho_{Y^k}$ and 
\begin{equation}\label{e:rho}
	\rho_{Y^k} = \sum_{j=0}^{2^k-1} p_j \qu{\phi_j}\dqu{\phi_j}	
\end{equation}
where the states $\qu{\phi_j}=\otimes_{i=1}^k\qu{\phi_{j_i}(Y_i)}$ and $j_i$ is the $i$-th bit of the binary representation of $j$. Note that $\rho_{Y^k} $ is a mixture of pure states weighted by the 
probabilities $p_j, \ j \in \{ 0, \ldots , 2^k-1 \}.$ Accordingly,  we write 
$p_j = \Pr[X^k = j] $ where $j$ is seen in its binary representation
(e.g., for $k=2, \ \ p_0 = \Pr[X^2 = 00], \ p_1 = \Pr[X^2= 01], \ldots $).
Observe that $S(\rho_{Y^k_1})=S(\rho_{Y^k_2})$ since there exist a unitary transformation $U$ such that $U\rho_{Y^k_1}U^{-1}=\rho_{Y^k_2}$.



We now establish a relationship between the probability vectors 
$\mathbf{p}_{X^k} $ and the lower bound given in Equation~(\ref{e:hs4}).

Denote by $\rho$  the uniform distribution, that is, 
$q_j = 1/2^k , \ j=0,\ldots , 2^k-1. $ In this section we shall verify that if 
the probability distribution of a block $X^k$ from the PRG, say $\mathbf{p} $, 
is near enough the distribution $\rho $,  for a block of size $k$, then the lower bound of 
~\eqref{e:hs4} will be kept significantly near of $k-kS^{\ast}$, which is the best one can achieve.

Let $\mathbf{\sigma}_{Y^k} $ be the density operator corresponding to a $k-$length block 
$X^k$ generated by a fair Bernoulli sequence given that the $k$-length block $Y^k$ is
known, that is
\begin{equation}\label{e:sigma}
	\mathbf{\sigma}_{Y^k} = \sum_{j=0}^{2^k - 1} q_j \qu{\phi_j}\dqu{\phi_j}.
\end{equation}

In this section we establish some results relating von Neumann entropy with the trace 
distance $D\left(\rho_{Y^k}, \mathbf{\sigma}_{Y^k} \right) $ between 
$\rho_{Y^k} $ and $\mathbf{\sigma}_{Y^k}$.  
Recall that the  trace distance between two quantum states 
$\mathbf{\rho} $ and $\mathbf{\sigma} $ is
defined by
\[
           D(\mathbf{\rho}, \mathbf{\sigma} ) = 
        \frac{1}{2}\mathrm{tr}\left| \mathbf{\rho} - \mathbf{\sigma} \right|
\]
where $|A| = \sqrt{A^{\dagger}A} $. We shall also need the trace distance between 
probability vectors, say $\mathbf{a} $ and $\mathbf{b}, $  defined by 
\[
          D(\mathbf{a}, \mathbf{b} ) = \frac{1}{2} \sum_{j} |a_j - b_j|.
\]

The trace distance can be used to measure how biased a probability distribution is 
compared to a fair Bernoulli sampling. Given a probability distribution $\mathbf{p}$, we 
call the {\em bias} of $\mathbf{p}$ the value $B(\mathbf{p})=D(\mathbf{p},\rho)$ 
where  $\rho$ is the uniform distribution. 
\begin{proposition}\label{prop:2}\em
Let $\varepsilon > 0 $ be an arbitrary real number. If
\begin{equation}
B\left( \mathbf{p} \right) \leq \varepsilon
\end{equation}
then 
\begin{equation}
D\left(\rho_{Y^k} , \mathbf{\sigma}_{Y^k} \right) \leq \varepsilon
\end{equation}
where $\rho_{Y^k} $ is the state defined in Equation~\eqref{e:rho}. 
\end{proposition}
\begin{proof}
Denote $\mathbf{\gamma}_j = \qu{\phi_j}\dqu{\phi_j} .$ From the strong convexity of the 
trace  distance we have:
\begin{eqnarray}
D\left(\sum_{j=0}^{2^k - 1} p_j\mathbf{\gamma}_j, 
           \sum_{j=0}^{2^k -1}\frac{1}{2^k} \mathbf{\gamma}_j \right)
& \leq & 
D\left( \mathbf{p}, \rho \right) 
+\sum_{j=0}^{2^k-1} p_j D\left(\mathbf{\gamma}_j,\mathbf{\gamma}_j\right)\\
&  = & D\left(\mathbf{p},\rho \right) 
\end{eqnarray}
which concludes the proof. \hfill $\square$
\end{proof} 
In the proof of the next proposition we shall apply Fannes' 
inequality (see \cite{bengtsson06} for more details about this equality):
\begin{equation}\label{e:fannes}
\left| S(\rho) - S(\sigma) \right|  \leq 2 D\left(\rho , \sigma \right)
                       \textrm{ln}\left(\frac{N}{2D(\rho , \sigma)} \right)
\end{equation}
where it is assumed that $D(\rho, \sigma ) \leq 1/(2\mathrm{e}) $ and $N$ is the 
dimension of the Hilbert space dimension where the states live in.
\begin{theorem}\label{thm:positivo}\em
If the conditions in Proposition~\ref{prop:2} hold, that is, if
$B(\mathbf{p} ) \leq \varepsilon $, then
\begin{eqnarray}
|H(X^k)-S\left(\rho_{Y^k}\right)-(H(X^k)-S\left(\mathbf{\sigma}_k\right))|&=&|S\left(\rho_{Y^k}\right) - S\left(\mathbf{\sigma}_k\right)|\\ & \leq & 
2\mathrm{ln}2 (k-1)\varepsilon  
+ 2\varepsilon \mathrm{ln}\frac{1}{\varepsilon}. \label{e:thm:positivo} 
\end{eqnarray}
\end{theorem} 
\begin{proof}
Observe that the function $- x\mathrm{ln}x $ is monotonous in the interval $(0,1/e)$. Therefore, 
assuming $0 \leq \varepsilon \leq 1/e $ and for $N=2^k $, we have: 
\begin{eqnarray}
| S\left(\rho_{Y^k}\right) - S\left(\mathbf{\sigma}_k\right) | 
&  \overset{(a)}{\leq} & 2D( \rho_{Y^k}, \mathbf{\sigma}_k )\textrm{ln}
\frac{2^k}{2D(\rho_{Y^k}, \mathbf{\sigma}_k ) }  \nonumber\\ 
& \overset{(b)}{=} &
2\mathrm{ln}2 (k-1)D(\rho_{Y^k}, \sigma_k ) 
        + 2 D(\rho_{Y^k} , \mathbf{\sigma}_k )\textrm{ln}
          \frac{1}{D(\rho_{Y^k}, \mathbf{\sigma}_k )} \\ 
& \overset{(c)}{\leq} &
2\mathrm{ln}2\  (k-1)\varepsilon + 2 \varepsilon \mathrm{ln}\frac{1}{\varepsilon} 
\end{eqnarray}
where $(a)$ results from Fannes' inequality, $(b)$ is due to logarithm properties and, 
$(c) $ is due to Proposition~\ref{prop:2}. $\square $
\end{proof}
This result states that if a PRG is such that the probability distribution of
its output $X^k ,$ say, $\mathbf{p} $ (possibly conditioned on the past), is  near enough
 the fair distribution $\rho$, then Eve's  uncertainty is kept near the 
maximum $H\left(X^k|Y^k,Z^k\right) = k-kS^{\ast} $ (see Equation~\eqref{e:hs2b}). 

Note that the distribution of $\mathbf{p} $ is induced by  
the random secret seed of the PRG, $X^{(n)} $,  which is chosen with uniform distribution. Consequently, any 
practical use of Equation~(\ref{e:thm:positivo}) will depend on the Eve's capability to 
estimate that distribution and clearly, on the PRG being used. 
For instance, suppose we want to upper bound the right side of~\eqref{e:thm:positivo} with 
a given \textit{tolerance} defined by a  positive  real number $\delta $. After some 
simple algebraic manipulation we obtain that
\begin{equation}\label{e:limitek}
k < 1 + \frac{1}{2\mathrm{ln} 2}\left(\frac{\delta}{\varepsilon}\right) - 
\frac{\mathrm{ln}\left(\frac{1}{\varepsilon}\right)}{\mathrm{ln} 2} . 
\end{equation}
For the case of $\varepsilon=\delta$ we get the simple bound
\begin{equation}\label{e:limitek2}
k < 1 + \frac{1}{2\mathrm{ln} 2} - 
\frac{\mathrm{ln}\left(\frac{1}{\varepsilon}\right)}{\mathrm{ln} 2} \leq 
 1 + \frac{1}{2\mathrm{ln} 2} + 
\frac{\left(\frac{1}{\varepsilon}\right)-1}{\mathrm{ln} 2}\leq 1+ \frac{1}{\varepsilon \mathrm{ln} 4}. 
\end{equation}

Additionally, when  the conditions of Proposition~\ref{prop:2} hold, that is, for 
bias $B(\mathbf{p}) < \varepsilon $, we can rewrite~(\ref{e:limitek}) as 
\begin{equation}\label{e:limitekB}
k < 1 + \frac{1}{2\mathrm{ln}2}\left(\frac{\delta }{B(\mathbf{p})}\right) - 
           \frac{\mathrm{ln}\left(1/B(\mathbf{p}\right))}{\mathrm{ln} 2}.
\end{equation}
Note that the right-hand side of \eqref{e:limitekB} approximates  
$\frac{1}{2\mathrm{ln}2}\left(\frac{\delta }{B\left(\mathbf{p}\right)}\right) $ as 
$B(\mathbf{p}) $ tends to  zero.  
In detail, Equation~(\ref{e:limitekB}), at the light of  Theorem~\ref{thm:positivo}, provides a way to compute the largest block whose  uncertainty remains near $k-kS^{\ast}$ (up to $\varepsilon$), given an upper bound of the bias of $\mathbf{p}$. However, 
a word of advice is necessary: from  its very definition, $\mathbf{p} $ 
depends on $k$  and also on the position $i$ the block start, because  $X^k = X_{i+1},X_{i+2},\ldots , X_{i+k} $. 
So, the use of Equation~\eqref{e:limitekB} to establish a bound of a secure block relies on a bias difficult to compute for standard PRG's. 

The following corollary clarifies the meaning of Theorem~\ref{thm:positivo} from 
an asymptotic point of view. 
\begin{corollary}\label{corol:thmpositivo}\em
Given a PRG, let $\mathbf{p}$ be the probability distribution of a 
$k$-length generated block, and let $f(k,n)$ and $g(n)$ positive functions such that:
 \begin{itemize}
 \item $\lim_{n\to \infty} g(n)=+\infty$
 \item $\lim_{n\to \infty} g(n) f(g(n),n)=0$. 
 \end{itemize}
Then, if $B(\mathbf{p}_{PRG})\leq f(k,n)$ and $k\leq g(n)$,
$$ 
\lim_{n\to \infty} 
|S\left(\rho_{Y^{g(n)}}\right) - S\left(\mathbf{\sigma}_{g(n)}\right) |=0.
$$ 
\end{corollary}

We now discuss the results above. The idea is that $n$ is the size of the seed of the $PRG$ and $k$ is the size of the block. If one chooses $k\leq g(n)$ for some $g$ and the bias of the PRG is smaller the $f(k,n)$ for some $f$ fulfilling the conditions of Corollary~\ref{corol:thmpositivo}, then the information Eve can retrieve from blocks of size $g(n)$ is as close to the ideal case as desired, just be choosing a larger $n$. A good PRG is one for which $n<<g(n)$, so that the block size could be larger than the seed and still, little information about the seed is revealed.

In the next section we make a comparison between classical $\mathtt{XOR}$ and quantum 
$\mathtt{QC}$ Brassard's schemes for authentication of classical messages.

\section{\label{sec:comparison} Improving Key-Tag Secrecy }
\begin{figure}[htbp]
\centerline{\epsfig{file=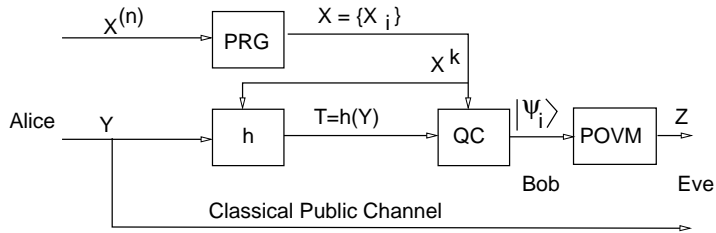,scale=0.5}}
\caption{Authentication scheme with a single key $X^{(n)}$  }
\label{f:qauthitsugestion}
\end{figure}

In the last section we compared Eve's equivocation on $X$ for the $\texttt{XOR}$ 
and $\texttt{QC}$ schemes when she has access both to the message $Y$ and its quantum 
encoded version, which she observes from the quantum channel. We concluded that the 
equivocation is kept above some lower bound depending on the quality of the PRG.
In this section we include a hash function $h$ in the scheme 
(see Figure~\ref{f:qauthitsugestion}) in such a way that Eve only accesses the public 
message $Y$ and the quantum encoded version of the tag $T=h(Y)$.
Thanks to that modification we shall demonstrate that is feasible 
to improve the secrecy of the key and of the tag simultaneously.


By information-theoretic secrecy, as usually, we mean $I(W; V) = 0 +O(2^{-n}) $ or 
equivalently, the equivocation $H(W|V) = H(W)- O(2^{-n}) $, where $W$ is the secret to be protected and $V$  
is the piece of data available to the eavesdropper. Our derivations will focus in the equivocation $H(W\mid V)$ to measure the quality of the scheme. 
Then, the information to be protected is $W = \left( T, X^k \right) $ and the information available, from Eve's viewpoint, is $V = \left( Y, Z \right)$. We
investigate the uncertainty of the tag  $H(T\mid Y,Z) $ and the uncertainty of the
key $H(X^k\mid Y,Z) $. 
We  assume that $X^k$, is independent of the message  $Y$ and that the hash function is selected 
from the $\varepsilon-$almost universal-2 class of hash functions,
which we refer in the following just as \textit{hash functions}. 


\subsection*{Modified classical case}

Consider a simple modified setup  
where a $\mathtt{XOR}$ gate is taken in place of the $\texttt{QC} $ block in the scheme displayed in 
Figure~\ref{f:qauthitsugestion}. 

If  $\{X_i\} $ is a fair Bernoulli and a $k-$block of bits such that
$ k = \max \{ \lceil \log |T | \rceil , \lceil \log |\mathcal{H}| \rceil\ \} $
it is utilized per message,  then the scheme turns to be equivalent to the Wegman-Carter 
one-time pad scheme. Indeed, in this situation $h$ is in fact drawn uniformly from $\mathcal{H}$, then 
\begin{eqnarray}
H\left(T, X^k| Y, Z^k\right) & \overset{(a)}{=} & H(T|Y, Z^k) + H(X^k| T,Y,Z^k)  \\
     & \overset{(b)}{=} & H(T| Y, Z^k ) \\  
     & \overset{(c)}{=} & H\left( T |Y \right) \\ 
     & \overset{(d)}{=} & \log \mid \mathcal{T}\mid . 
\end{eqnarray}
Where equality $(a)$ is due to chain rule for Shannon entropy, 
$(b)$ is due to the fact  that in the  classical setup $X^k = T\oplus Z^k $. 
The equality (c) is harder to obtain, indeed it follows from the properties of the $\varepsilon-$almost universal-2 class of hash function. Note that $T=h_{X^k}(Y)$ has a uniform distribution. Moreover $T|_{x_k}=h_{x^k}(Y)$ has also uniform distribution, and therefore, $T$ is independent of $X^k$. Since $Z^k=f(X^k,Y)$ we have that $ H(T| Y, Z^k )= H(T| Y)$. Equality $(d)$ is also due to the properties of hash functions. 
On the other hand, if $\{ X_i \} $ comes from a PRG, the Eve's uncertainty on 
the tag can,  eventually, decreases by observing the random variable $Z^k$. Indeed,
in general, $ H\left( T |Y, Z^k  \right) < H\left(T | Y\right) $.
Consequently, unconditional secrecy relative to $T$,  $H(T|Y,Z^k) = \log{|\mathcal{T}|}$ 
cannot be assured.  
 
\subsection*{Uncertainty of the tag in the quantum case}
In this subsection we introduce a condition to attain unconditional security of the tag in terms
of conditioned mutual information between $T$ and the $k-$block of bits of the key.
\begin{proposition}\label{prop:sufficient}\em
If $ I\left(T; X^k | Y, Z^k \right) = H(T) $ then the tag is
secure in the information theoretical sense, that is, $H(T|Y, Z^k )=H(T)$.
\end{proposition}
\begin{proof}

From the standard chain rule of Shannon entropy we have:
\begin{eqnarray}
H\left(T, X^k \mid Y, Z^k \right) & = & 
H\left(X^k \mid Y, Z^k\right)  + H\left(T \mid X^k, Y, Z^k\right)  \label{e:1}\\ 
& = & 
H\left(T | Y, Z^k \right)  + H\left(X^k | T, Y, Z^k \right). ~\label{e:2}
\end{eqnarray}
Then, comparing~(\ref{e:1}) and ~(\ref{e:2}) we obtain
\begin{eqnarray}
H\left( T | Y, Z^k \right) & \overset{(a)}{=} & 
                        H\left(T | X^k ,Y, Z^k \right) + \nonumber \\
& & 
H\left(X^k|Y, Z^k \right) 
- H\left(X^k|T, Y, Z^k\right) ~\label{e:3} \\
& \overset{(b)}{=} & 
H\left(T | X^k ,Y, Z^k \right) + 
I\left(T; X^k | Y, Z^k\right) ~\label{e:4} \\
& \overset{(c)}{=} & 
I\left(T; X^k | Y, Z^k \right) ~\label{e:5} 
\end{eqnarray}
Where $(a)$ is due to a simple manipulation of~\eqref{e:1} and~\eqref{e:2}, $(b)$ is 
definition of mutual information and $(c)$ follows because the hash function is determined
by $X^k$ and so, then $T = h(Y) $ is immediately calculated. That is,   
$H(T|X^k, Y, Z^k) = H(T|X^k,Y,T = h(Y)) = 0.$  
The results follows from~\eqref{e:5}. $\square $
\end{proof}
Eq.~\eqref{e:5} clearly indicates that in order to increase Eve's uncertainty 
about $T$ we must maximize the mutual information between the block $X^k$ 
and the tag $T.$ This is the \textit{information-theoretical} hint that motivates 
the scheme  presented in Figure~\ref{f:qauthitsugestion}. Note that in this case we make the tag $T$ depend of $X^k$, increasing thus their mutual information. In Brassard scheme (see Figure~\ref{f:qauthbrassard}) the hash function is fixed in the beginning, and therefore $I\left(T; X^k |V' \right)=0$ where $V'$ is the observation that Eve can perform in Brassard's scheme.

It is remarkable to be possible to attain unconditional security of the tag
using non-fair Bernoulli for $X$ with the proposed of Figure~\ref{f:qauthitsugestion}. This fact is in sharp
contrast with the classical setup for which only Bernoulli sequences can assure that requirement.

Thus, a good approximation is to use PRG for the sequence of $X$, and the mutual information $I(T;X^K|Y,Z^k)$ is as high as the PRG is unbiased, since that mutual information is mediated by 
the random variable $Z^k$.


It is clear that, if we are dealing with real PRGs (that do not generate a  sequence of fair Bernoullis), then the conditions of 
Theorem~\ref{thm:positivo} should be considered in order to evaluate the number 
of messages that can  be authenticated before leaking too much information.
Another possibility to apply the scheme of Figure~\ref{f:qauthitsugestion} is to spend just 
$k=\log |\mathcal{T}|$ key bits per message to protect the current tag. This approach is similar 
to Brassard's scheme, but improves it since the tag is protected by the quantum coding. Observe that as 
$ \log | \mathcal{T} | < \log | \mathcal{H}|  $, this scheme is less costly in terms of key 
consumption. 

\subsection*{Uncertainty of the key in the quantum case}

In this case, the bounds derived in Section~\ref{sec:comparing} remain valid, namely the inequality~\eqref{e:hs4} that we recall
\begin{equation}\label{e:hs4b} 
H\left( X^k|Y^k, Z^k \right)  \geq H(X^k)-  S(\rho_{Y^k} ) = 
H(X^k)-H(\lambda ). 
\end{equation}
In this case, since the measurement $Z^k$ is on the quantum encoding of the tag, and not on the quantum encoding of $Y$, the uncertainty is greater than that of the case discussed in Section~\ref{sec:comparing}.

So, with the scheme of Figure~\ref{f:qauthitsugestion}, not only we obtain a high equivocation about the tag, but we also increase the uncertainty of the sequence $X^k$ and, therefore, also of the seed $X^{(n)}$ of the PRG. Observe that Theorem~\ref{thm:positivo} and inequality~\ref{e:limitek} are also valid for this scheme, and can be used to get bounds about the size of $k$ for which a threshold of information is leaked to Eve.

\section{Summary}
In this work we have investigated how quantum resources can improve the security of Brassard's classical
 message authentication protocol. We have started by showing that a quantum coding of secret bits offers
 more security than the classical \texttt{XOR} function introduced by Brassard. Then, we have used this
 quantum coding to propose a quantum-enhanced protocol to authenticate classical messages, with improved
 security with respect to the classical scheme introduced by Brassard in 1983. Our protocol is also more
 practical in the sense that it requires a shorter key than the classical scheme by using the pseudorandom
 bits to choose the hash function. We then establish the relationship between the bias of a PRG and
the amount of information about the key that the attacker can retrieve from a block of authenticated
 messages. Finally, we prove that quantum resources can improve both the secrecy of the key generated by
the PRG and the secrecy of the tag obtained with a hidden hash function.

\section*{Acknowledgments}
F. M. Assis acknowledges partial support from Brazilian National 
Council for Scientific and Technological Development (CNPq) under Grants No.~302499/2003-2 
and CAPES-GRICES No.~160.
P. Mateus and Y. Omar thank the support from project IT-QuantTel, as well as from Funda\c{c}\~{a}o para a Ci\^{e}ncia e a
Tecnologia (Portugal), na\-mely through pro\-grams POC\-TI/PO\-CI/PT\-DC and proj\-ects 
PTDC/EIA/\-67661/2006 QSec and  PTDC/EEA-TEL/103402/2008 QuantPrivTel, partially funded by FEDER (EU).


\begin{thebibliography}{10}
\bibitem{Bras:83}
G.~Brassard,
\newblock ``On computationally secure authentication tags requiring short secret
  shared keys'',
\newblock in {\em Advances in Cryptology} (Springer-Verlag, 1983), pp. 79--86.

\bibitem{WeCa:81}
M.~N. Wegman and J.~L. Carter,
\newblock ``New hash functions and their use in authentication and set equality.''
\newblock {\em J. Comput. Syst. Sci.}, 22:265--279 (1981).

\bibitem{BM84}
M.~Blum and S.~Micali,
\newblock ``How to generate cryptographically strong sequences of pseudo-random  bits'',
\newblock {\em SIAM J. on Computing}, 13(4):850--864 (1984).

\bibitem{NC:2000}
M.~A. Nielsen and I.~L. Chuang,
\newblock {\em Quantum Computation and Quantum Information}
\newblock (Cambridge University Press, New York, 2000).

\bibitem{WSta:98}
W.~Stallings,
\newblock {\em Cryptography and Network Security: Principles and Practice}
\newblock (Prentice-Hall, Inc., New Jersey, 2nd edition, 1998).

\bibitem{larsson08.1}
J. Cederl\"{o}rf and J. Larsson,
\newblock `` Security Aspects of the Authentication Used in Quantum Cryptography '',
\newblock {\em IEEE Transactions on Information Theory}, 54(4):1735--1741
           (2008).

\bibitem{goldreich99}
O.~Goldreich,
\newblock {\em Modern Cryptography, Probabilistic Proofs and Pseudorandomness},
\newblock (Springer, 1999)

\bibitem{sidok05.1}
A.~Sidorenko and B.~Shoenmakers,
\newblock ``State Recovery Attacks on Pseudorandom Generators'',
\newblock in {\em Western European Workshop on Research in Cryptology, Lectures Notes in 
Informatics (LNI)}, 74:53--63 (2005).

\bibitem{Spi:96}
T.~P. Spiller,
\newblock ``Quantum information processing: cryptography, computation, and
teleportation'',
\newblock {\em Proceedings of the IEEE}, 84(12):1719--1746 (1996).

\bibitem{BS:98}
C.~H. Bennett and P.~W. Shor,
\newblock ``Quantum information theory'',
\newblock {\em IEEE Transactions on Information Theory},  44(6):2724--2755
 (1998).

\bibitem{BB84:84}
C.~H. Bennett and G.~Brassard,
\newblock ``Quantum cryptography: public-key distribution and coin tossing'',
\newblock  in {\em Proceedings of IEEE International Conference on Computers,
Systems, and Signal Processing, Bangalore, India, 1984} 
 (IEEE Press, New York, 1984), pp. 175--179.

\bibitem{Paris:04}
M.~G.~A. Paris and J.~Reh\'{a}cek,
\newblock {\em Lectures Notes in Physics, Quantum State Estimation }
Springer, Berlin, 2004

\bibitem{Shor:94}
P.~W. Shor,
\newblock ``Algorithms for quantum computation: Discrete logarithms and
  factoring'',
\newblock in {\em {IEEE} Symposium on Foundations of Computer Science}, pp.
  124--134, (1994).

\bibitem{bengtsson06}
I. Bengtsson {and} K. Zyczkowski,
\newblock {\em Geometry of Quantum States}
\newblock (Cambridge University Press, 2006).

\bibitem{Damgaard:04}
I. Damgaard, T. Pedersen, L. Salvail,
\newblock quant-ph/0407066 (2004).
 
\end{thebibliography}
\end{document}